# Multi-species breath biomarker profiling with an ultra-broadband (2.9–11.5 μm) spectroscopic platform

Roderik Krebbers,[1] Marleen Huisman,[1] Kees van Kempen,[1] Joris Meurs,[1] Amir Khodabakhsh,[1,2] Simona M. Cristescu[1*]

1. Trace Detection Laboratory, Department of Spectroscopy and Catalysis, Institute for Molecules and Materials, Radboud University, Heyendaalseweg 135, 6525 AJ Nijmegen, The Netherlands
2. Presently at: Faculty of Science and Engineering, Maastricht University, Paul Henri Spaaklaan 1, 6229 GS, Maastricht, The Netherlands
*simona.cristescu@ru.nl

**Abstract**
Online, comprehensive molecular profiling of exhaled breath provides a non-invasive window into human metabolism, yet current optical platforms are restricted by narrow instantaneous spectral coverage. Here, we present a novel ultra-broadband mid-infrared spectroscopic platform that enables simultaneous, high-sensitivity detection of a comprehensive profile of breath biomarkers. By integrating an intrapulse difference-frequency generation (IDFG) supercontinuum source spanning 2.9–11.5 μm (2580 $cm^{-1}$) with a custom-built Fourier transform spectrometer, we achieve a spectral resolution of 0.1 $cm^{-1}$ – surpassing current laser-based approaches. Combined with a standardized online sampling system, the platform achieves sensitivities in the tens of parts per billion over three minutes, resolving dynamic metabolic changes of ammonia, methane, isoprene, acetone, carbon monoxide, and nitrous oxide. We demonstrate the system's utility through proof-of-concept case studies tracking responses to fasting, protein intake, and smoking. This calibration-free platform establishes a powerful and versatile tool for online breath analysis, with broad potential in clinical diagnostics and exposure monitoring.

## Introduction

Breathomics is a rapidly advancing field in medicine that provides a snapshot of the body's current biochemistry by detecting volatile biomarkers in exhaled breath. These biomarkers arise as (by)products of endogenous metabolic processes occurring throughout the body and carry valuable physiological and disease-associated insights: an overview of typical compounds and their established physiological basis is provided in Table 1. The ability to non-invasively detect changes in biomarker concentrations in (nearly) real-time makes breathomics powerful in pinpointing specific biochemical pathways involved in a health condition and revealing their instant alterations. This provides the physician with a rapid and accessible tool to facilitate faster decision-making in screening, diagnosis, and treatment, ultimately leading to more timely and effective interventions.

Advanced technologies based on mass spectrometry, ion mobility spectrometry, or laser spectroscopy are currently employed for the detection of specific compounds or untargeted breath profiles [1–3]. Laser-based optical spectroscopy is preferred for breath analysis targeting a single molecular species [4,5]. It provides fast measurements with high selectivity and sensitivity. Most of the FDA-approved breath tests rely on the detection of a single compound as an indicator for a certain health condition, such as nitric oxide, a well-established biomarker of airway inflammation, particularly in asthma [6]. Importantly, these compounds have known biochemical pathways and, with a few exceptions, are typically well-detectable and quantifiable with optical methods. The clinical practice, however, might demand a broader characterization of health conditions or diseases and thus a demand for multiple breath metabolites measured simultaneously.

Therefore, versatile mass spectrometric (MS) systems such as SIFT-MS [7], PTR-ToF-MS [7,8], and GC-MS [9] are widely used within breath research. Currently, mass spectrometry-based methods stand as the current golden standard in (un)targeted breath analysis in clinical studies [5,10]. Whilst GC-MS provides very good sensitivity, it inherently samples offline, requiring long processing times and lacking the temporal resolution needed to monitor rapid metabolic changes. In comparison, SIFT-MS and PTR-ToF-MS can be used in online settings for real-time analysis of exhaled breath [7,8]. Identifying isomeric compounds poses a challenge for MS techniques, and quantifying their concentrations often requires reference and calibration measurements. Finally, detecting small molecular compounds in breath, such as nitric oxide (NO), carbon monoxide (CO), methane ($CH_4$), hydrogen sulfide ($H_2S$), ethylene ($C_2H_4$), or ammonia ($NH_3$) poses an additional challenge to these techniques.

The introduction of broadband infrared laser-based spectroscopic methods offers a practical solution that combines the advantages of MS-based methods and laser-based spectroscopy. The ultrabroadband spectral coverage of emerging supercontinuum sources enables the versatility and flexibility of detecting many molecular compounds simultaneously in a single device, similar to MS systems. Moreover, the high sensitivity, unambiguous identification, and quantification of molecular compounds without requiring calibration procedures, which are specific to laser-based spectroscopy, remain.

Breath analysis with broadband infrared spectroscopy has been demonstrated using thermal lamps for the broadest spectral coverage (3500 $cm^{-1}$), at the cost of limited sensitivity [11]. Even state-of-the-art mid-infrared frequency-comb systems, while offering high resolution, typically provide limited instantaneous coverage, with the broadest spectral coverage reported being 1010 $cm^{-1}$ [12]. It should be noted that the latter coverage was obtained by combining separate measurements of two different laser systems. While these studies demonstrated spectroscopic measurements of breath compounds, the setups used are still optical laboratory-bound and lack online measurement capabilities or advanced breath sampling capabilities, such as the use of capnography and heated sampling lines. Moreover, beyond the spectroscopic capabilities of these instruments, breath analysis methods require standardization of sampling and analysis to ensure reproducibility across different instruments, as was demonstrated in the comparison of different MS-based methods [8,13,14].

Here, we report an ultra-broadband, online spectroscopic platform that overcomes these limitations by spanning a 2.9–11.5 µm (2580 $cm^{-1}$) range in a single scan with 0.1 $cm^{-1}$ spectral resolution. By integrating a novel ultra-broadband intrapulse difference-frequency generation (IDFG) source with a custom-built Fourier transform spectrometer (FTS) [15] and a standardized breath sampling system, our platform enables the simultaneous quantification of multi-species biomarkers - ranging from light inorganic gases to hydrocarbons - without the need for laser tuning or external calibration. This integrated architecture is robust, transportable, and functional outside laboratory conditions [16]. Furthermore, we developed sophisticated spectroscopic analyses to fully leverage the high sensitivity and broad spectral density of the acquired data [17]. The fully integrated platform with its three core systems – the dedicated breath sampling unit, the IDFG infrared supercontinuum source, and the FTS – is shown in Fig. 1.

We demonstrate that the system simultaneously captures an unprecedented spectrum of biomarkers across diverse molecular classes in exhaled breath associated with established biochemical pathways, including acetone, isoprene, nitrous oxide ($N_2O$), $NH_3$, CO, and $CH_4$.

Using a standardized online sampling methodology in which we sample end-tidal breath samples directly into our system, we resolve dynamic metabolic changes in response to interventions such as fasting, protein intake, and smoking. These case studies produce distinct, detectable shifts in metabolic signatures, representing a significant leap from laboratory-scale laser demonstrations to a robust modality for comprehensive, non-invasive molecular profiling. Ultimately, this first-of-its-kind laser-based platform combines sensitivity and selectivity with broad applicability, offering strong potential for clinical translation and personalized medicine.

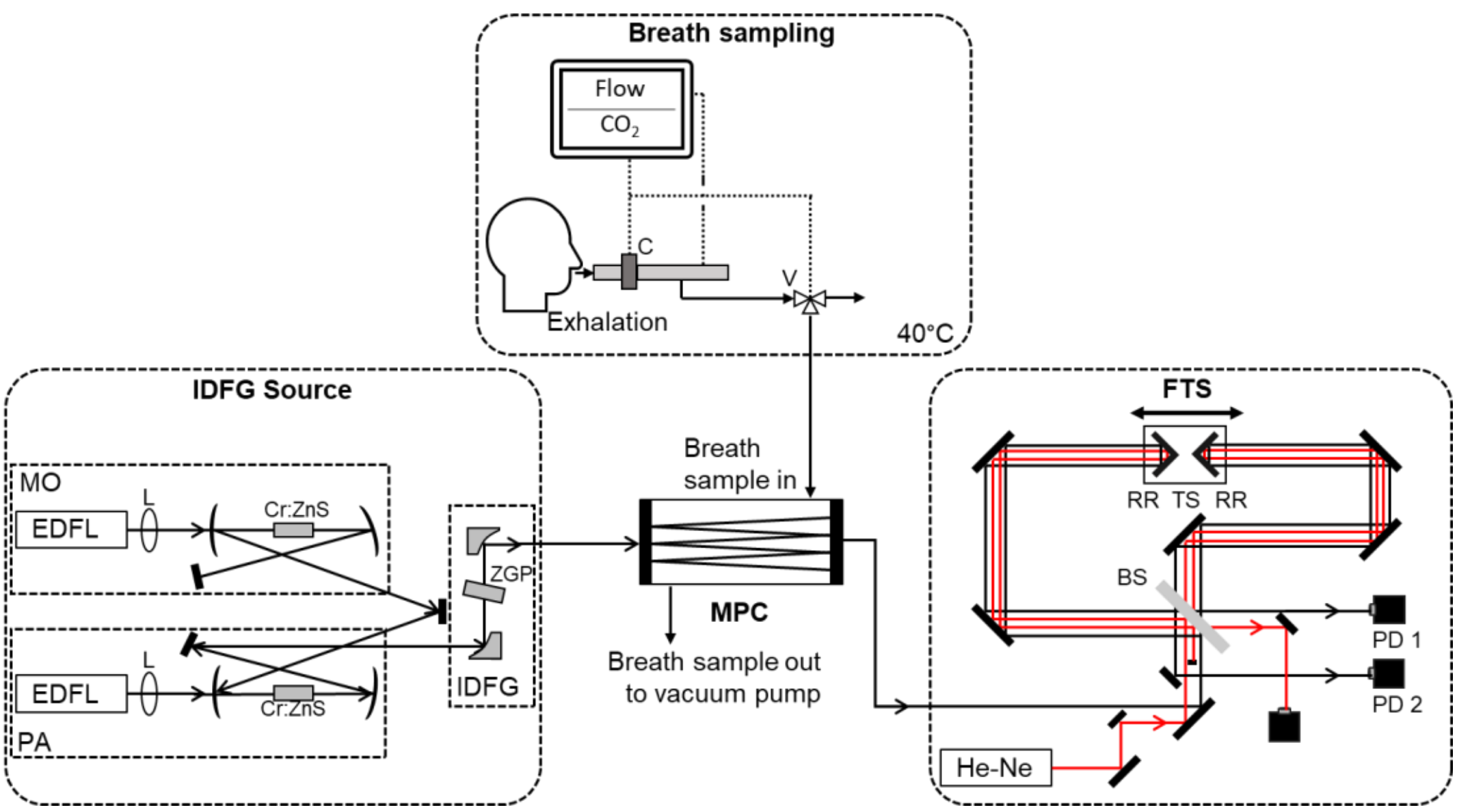

***Fig. 1: System design of the spectroscopic breath platform.*** *The integrated platform with its three core systems: breath sampling system, laser source, and spectrometer. EDFL: erbium-doped fiber laser, L: lens, MO: master oscillator, PA: power amplifier, IDFG: intrapulse difference-frequency generation, Cr:ZnS: chromium-doped zinc selenide crystal, ZGP: zinc germanium phosphide crystal, C: Capnograph, V: three-way valve, MPC: multi-pass cell, FTS: Fourier transform spectrometer, He-Ne: helium-neon laser, BS: beam splitter, TS: translation stage, RR: retroreflector, PD: photodetector.*

## Results

A central advantage of the information-rich, broadband, high-resolution spectra provided by our system is the ability to perform both untargeted – using the entire spectrum – and targeted – focusing on selected compounds – analyses within the same dataset. This dual approach combines the discovery power of global spectral profiling with the precision of compound-specific quantification, offering a flexible analytical framework for breathomics.

### *Untargeted spectral mapping*

To map the comprehensive spectral signature of exhaled breath, we first performed a principal component analysis (PCA) on the raw absorbance data. This approach utilizes the information-dense spectra to extract intervention-related changes in the breath composition without prior selection of compounds.

As shown in Fig. 2.a, the scores plot reveals a clear clustering of breath samples according to the intervention, with the first two principal components (PC1 and PC2) successfully resolving the distinct metabolic signatures of fasting, protein intake, and smoking. The high reproducibility of the standardized sampling system is evidenced by the tight clustering within the 95% confidence intervals across different participants.

The corresponding loadings (Fig. 2.b) highlight the broad spectral regions driving this separation. Notably, a close examination of the PC2 loadings (Fig. 2.c) reveals features that match the rotational-vibrational structure of $CH_4$, as confirmed by HITRAN2020 simulations [18].

This direct correlation between the utilized variance and specific molecular features validates the platform's ability to retrieve biologically relevant information from complex spectral data.

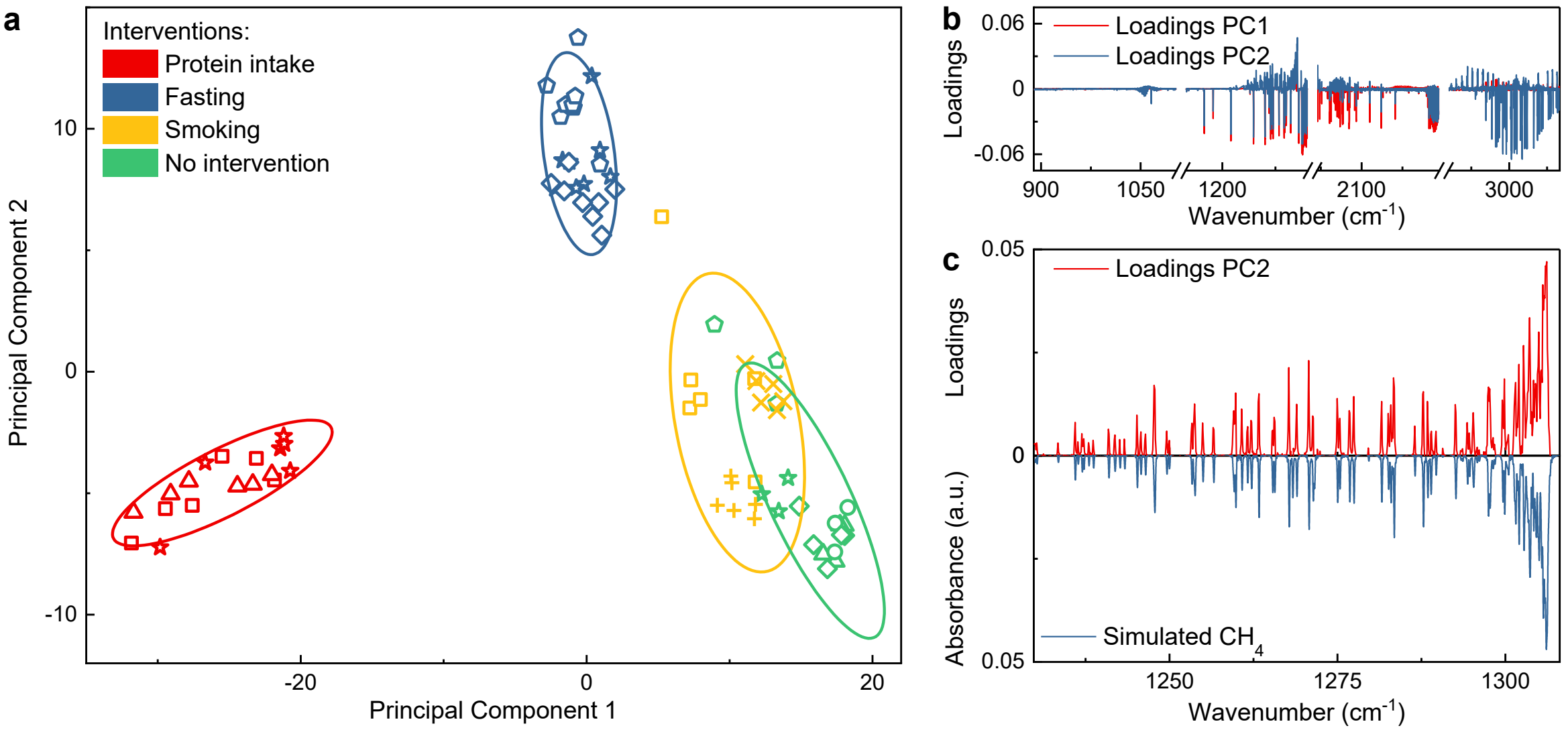

***Fig. 2: Principal component analysis (PCA) of the breath samples of different interventions.*** *(**a**) PCA scores plot of the absorbance spectra of all breath samples, with associated intervention indicated. The ellipse indicates the 95 % confidence interval of the different interventions. Symbols denote individual participants. (**b**) Loadings corresponding to PC1 and PC2. (**c**) A zoom to a section of the PC2 loadings shows its match to $CH_4$ features (simulated with HITRAN, shown inverted).*

### *Targeted retrieval and high-resolution quantification*

Following the untargeted mapping, we performed targeted retrieval to quantify specific biomarkers across the full 2.9–11.5 µm spectral range. Utilizing the high 0.1 $cm^{-1}$ spectral resolution of the IDFG-FTS platform, we successfully resolved the individual rotational-vibrational lines of various compounds, including acetone, isoprene, $NH_3$, CO, $N_2O$, and $CH_4$ (Fig. 3). As a result, these molecules could be fitted simultaneously without spectral crosstalk.

Quantitative analysis was performed using a calibration-free approach based on the Beer-Lambert law, where concentrations were retrieved by fitting simulated reference absorbance spectra of compounds of interest (from HITRAN2020 [18] and the PNNL database [19]) to the measured broadband absorbance spectra (Fig. 3.b-d).

The breath data samples are presented as both absolute absorbance and relative absorbance spectra. Absolute absorbance spectra – measured against a dry nitrogen background – allow for the determination of the absolute concentrations. In contrast, relative absorbance spectra represent the difference between two breath samples. In this representation, absorbance peaks of abundant molecular compounds with minimal concentration variation (such as $CO_2$ and $H_2O$ absorption) are largely suppressed, enhancing the visibility of intervention-relevant concentration changes of weakly-absorbing compounds.

Examples of the absolute (Fig. 3.b-d) and relative (Fig. 3.e-g) absorbance spectra illustrate how intervention-related changes in the acetone and ammonia concentration become more apparent in the relative absorbance spectra, as the absolute absorbance spectra are dominated by strong $CO_2$ and $H_2O$ absorption features. Nevertheless, absolute $CO_2$ and $H_2O$ concentrations are valuable for assessing and correcting potential inconsistencies in the breath sample collection. A common practice in breath analysis is to report compound concentrations relative to the $CO_2$ content of the sample [20]. Although the commercial sampler reliably captures end-tidal breath, occasional small deviations in measured $CO_2$ can occur, likely due to physiological or instrumental sampling variations. Because end-tidal $CO_2$ reaches a stable plateau and is highly reproducible, $CO_2$ normalization is widely used to reduce variability between breaths and across participants.

*Analytical performance and detection limits*

Fitting simulated reference spectra to the measured absorbance spectra provides a comprehensive overview of the retrieved concentrations for all detected gases in a single sample (Fig. 4). This standardized approach facilitates a broad mapping of the exhaled breath composition, enabling a robust analysis of its various molecular biomarkers.

Table 2 summarizes the detection limits achieved for the molecular species retrieved across the following case studies. By focusing on the strongest absorption features within the 2.9–11.5 μm spectral window, we demonstrate the large spectral and dynamic range of the platform. Detection limits are defined here as the precision of the retrieved concentrations in breath samples and are thus specific to the wavenumber ranges used for fitting individual compounds. For species present at very high concentrations, such as $CO_2$ and $H_2O$ (percent-level abundances), relatively weak absorption lines were selected to avoid saturation, resulting in detection limits in the high-ppm range. Most other compounds exhibit detection limits on the order of tens of ppb, sufficient for tracking clinically relevant markers such as acetone, $NH_3$, and CO. While further improvements are required for sub-ppb targets like ethylene, this performance provides a robust foundation for the physiological monitoring presented in the following sections.

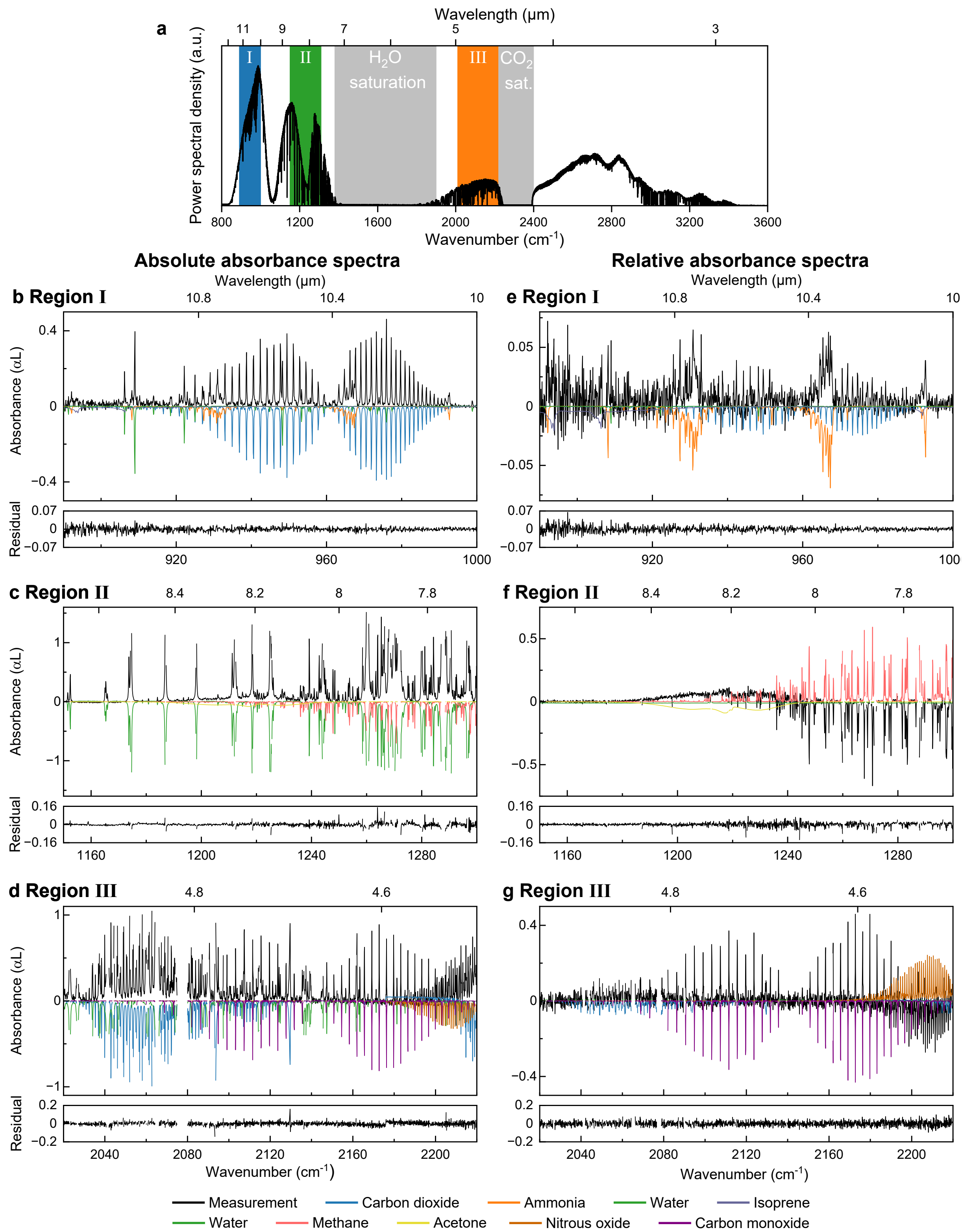

***Fig. 3. Power spectral density of the ultra-broadband source, and the absorbance spectra with fitted reference spectra. (a)*** *Power spectrum of a representative breath sample showing strong absorption features from water and carbon dioxide. Regions containing absorption of relevant molecular compounds are indicated (I, II, and III).* ***(b/c/d)*** *Measured absolute concentration absorbance spectra with fitted reference spectra and retrieved concentrations for regions I - III. Absolute concentration absorbance spectra are normalized to a blank, nitrogen gas sample.* ***(e/f/g)*** *Measured relative concentration absorbance spectra with fitted reference spectra and retrieved*

*concentrations for regions I, II, and III. Relative concentration absorbance spectra are normalized to a baseline breath sample from the same participant.*

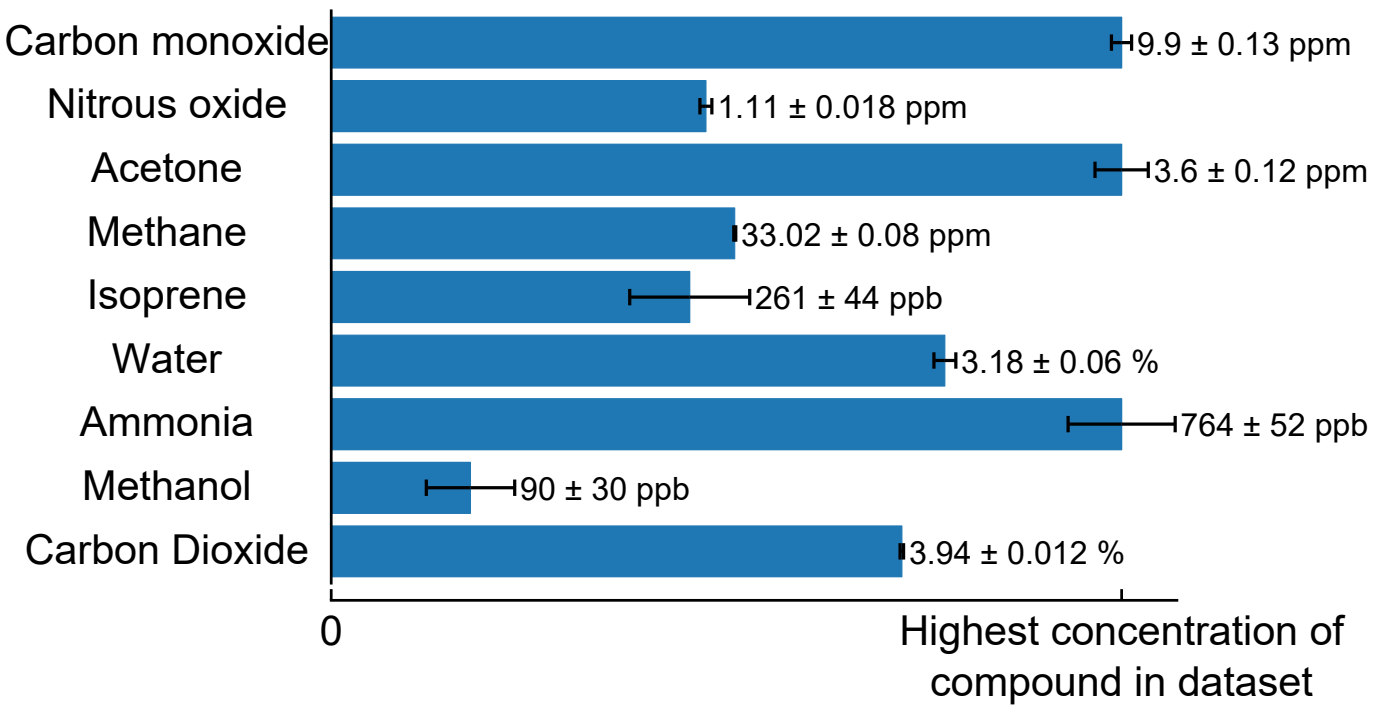

***Fig. 4. Retrieved concentrations of the spectra in Fig. 3.b-d.*** *Concentrations corresponding to the fits of the absolute absorbance spectra in Fig. 3.b-d. The bar length indicates the highest retrieved concentration of each compound across the 84 analyzed breath samples. Error bars represent the standard deviations between fits of the same sample.*

### *Tracking metabolic changes and exposure: intervention-based case studies*

To demonstrate the versatility of the broadband breath platform, we conducted three case studies targeting a wide range of molecular classes, ranging from inorganic compounds (e.g., $NH_3$), hydrocarbon compounds (e.g., $CH_4$, isoprene), and ketone compounds (e.g., acetone), to oxides (e.g., CO, $N_2O$) and alcohols (e.g., methanol). Importantly, all compounds were retrieved from a single 3-minute measurement without any modification of the instrument. For clarity, one compound is emphasized per case study (ammonia for protein intake, acetone for fasting, and carbon monoxide for smoking) while the full suite of simultaneously measured compounds is provided in the Supplementary Information.

### *Protein intake and ammonia monitoring*

In the first study, participants consumed a protein-rich shake, with breath samples collected hourly. As expected, exhaled ammonia levels increased, following amino acid catabolism [21–24] (Fig. 5.a). While these findings align with results from dedicated ammonia systems based on thermal oxidation [22], photoacoustic spectroscopy [25], or a selected ion flow tube (SIFT) mass spectrometry [21], our platform offers a distinct advantage: the simultaneous capture of the full molecular landscape. For example, the concurrent measurement of $CO_2$ allows for precise selection of the exhalation fraction (i.e. the end-tidal phase corresponding to $CO_2$ concentrations above 5 %) and supports normalization (details in Supplementary Information Fig. 1). This integration minimizes sampling variability and enhances the accuracy of ammonia quantification.

Furthermore, the broadband spectra revealed subject-specific metabolic signatures beyond the primary target. We observed a modest rise in isoprene, a byproduct of lipolytic cholesterol metabolism in skeletal muscle [26] (see Supplementary Information Fig. 2). These observations highlight the platform's capacity to concurrently capture metabolic pathways in a single acquisition, a key requirement for multi-species breathomics.

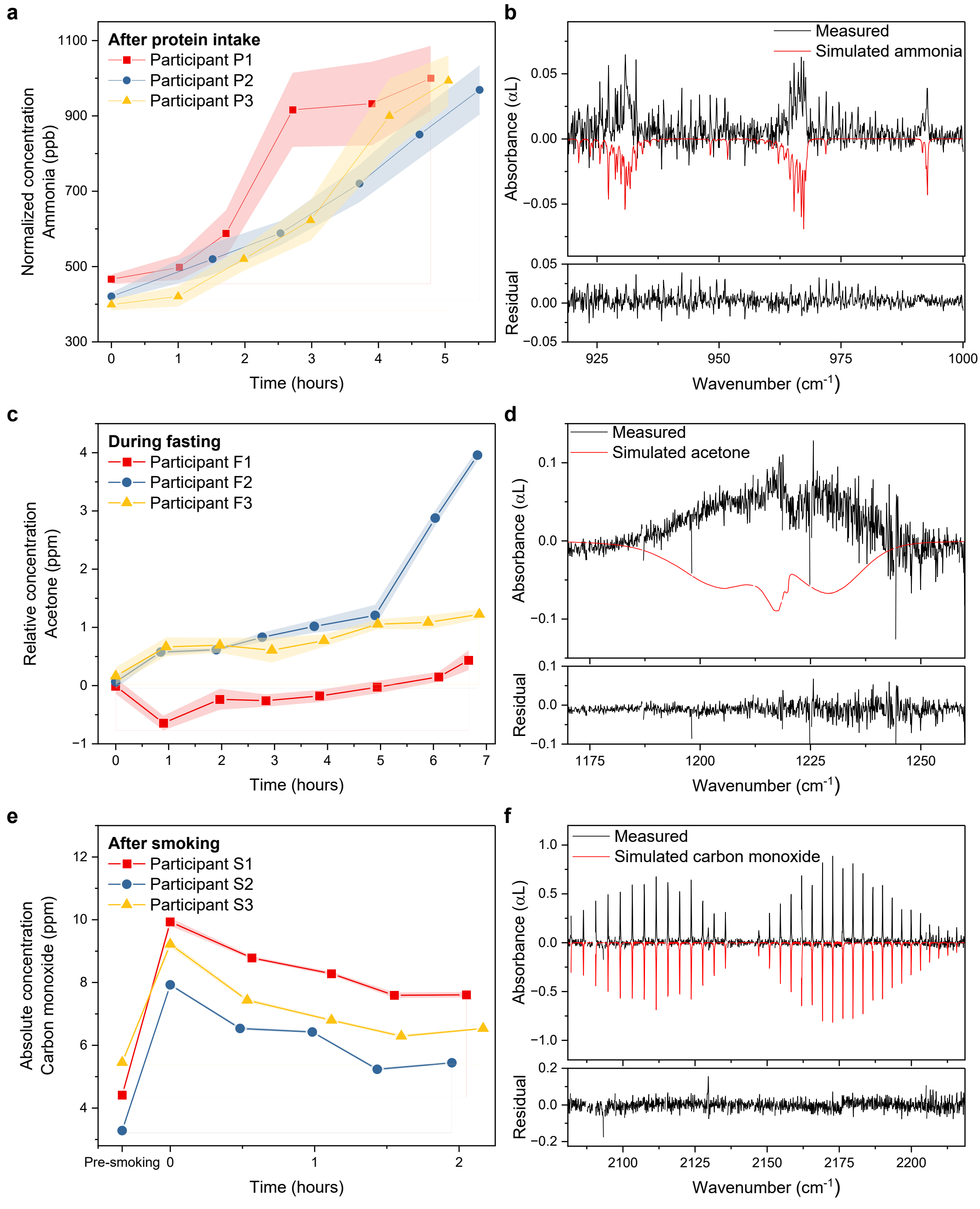

**Fig. 5. Dynamics of three biomarkers during case studies. (a)** Retrieved exhaled ammonia concentrations, showing an increase after protein intake at time = 0 h. Concentrations are normalized to the retrieved carbon dioxide levels (see Supplementary Information Fig. 1). **(b)** Measured (black) and simulated (from HITRAN) absorbance spectra of ammonia (in red, inverted), with the associated residual of the fit. (**c**) Retrieved relative change in acetone concentrations of participants during fasting. Measurements started 14 hours after fasting was initiated. (**d**) Measured (black) and reference (from PNNL, in red, inverted) absorbance spectra of acetone with the associated residual of the fit. (**e**) Retrieved absolute concentrations of CO showing a gradual washout over 2 hours following smoking a cigarette (at time = 0 h). (**f**) Measured (black) and simulated (from HITRAN, in red, inverted) absorbance spectra of CO with the associated residual of the fit. Shaded areas indicate measurement uncertainty, calculated as the standard deviation of five consecutive measurements of the same breath sample.

### *Fasting and acetone dynamics*

In a second case study, we monitored the metabolic transition during a period of 14 -22 hours of fasting. While traditional ketone monitoring requires blood or urine analysis, our platform enables the non-invasive tracking of these shifts with high temporal resolution. The acetone concentration was traced using the relative absorbance spectra (see Materials and Methods) to reduce the effect of the strongly overlapping high-absorbing water features. Consequently, the presented acetone concentration in Fig. 5.c refers to the change in acetone concentration (in ppb) compared to the participant's first measured breath sample (at time = 0 h). Exhaled acetone, a well-established marker of ketosis, fatty acid oxidation, and acetoacetate decarboxylation [21,27–30] showed a significant and progressive increase, as the body's metabolism adapts to maintain energy balance. Two participants exhibited a gradual increase in acetone concentration throughout the day, while one participant (F2) showed a pronounced spike toward the end of the measurement period, reflecting individual metabolic responses to fasting (Fig. 5.c). Interestingly, in this participant, the acetone spike coincided with an abrupt decrease in methane (see Supplementary Information Fig. 3). This apparent 'crossing point,' where acetone rises while methane falls, may reflect individual metabolic differences (e.g., energy substrate use or gut microbiota activity). The IDFG-FTS platform resolved these synchronized trajectories of methane and acetone at the individual level with high precision, demonstrating the system's sensitivity to subject-specific metabolic shifts during fasting (Fig. 5.c). While physiologically plausible, this observation remains exploratory and requires further study to confirm its potential as a biomarker of metabolic transitions during fasting.

To benchmark the quantitative accuracy of our platform, we compared the retrieved acetone trends with those measured by proton-transfer-reaction time-of-flight mass spectrometry (PTR-TOF-MS). We found an excellent correlation (R-squared value of 0.93), confirming the agreement between the two modalities (Supplementary Fig. 5). Notably, while the PTR-TOF-MS requires external calibration with gas standards, our platform achieved this high level of agreement using a calibration-free spectroscopic model, demonstrating its reliability for absolute quantification.

*Smoking and exogenous CO*

In the third case study, we investigated the system's capability for exposure monitoring by tracking the effects of cigarette smoking. Carbon monoxide (CO), a primary marker of combustion and tobacco smoke inhalation [31], was found to be already elevated (>2 ppm) pre-smoking, highlighting the prolonged systemic retention of this compound in the lungs. Upon smoking, a rapid elevation in exhaled CO was observed, followed by a gradual wash-out phase (Fig. 5.e). This decay is consistent with the known half-life of carboxyhemoglobin, which persists for several hours [31]. This case study demonstrates the platform's high dynamic range: while CO shows a clear smoking-related washout, the platform is able to accurately track other compounds simultaneously, even when their concentrations vary significantly (Supplementary Information Fig. 4).

## Discussion

In this work, we present the first fully functional platform that establishes ultra-broadband laser-based infrared spectroscopy as a powerful new modality for exhaled breath analysis. By overcoming the limitations of conventional single-frequency or narrowband approaches, which typically target a single or a few molecular compounds, the platform yields a comprehensive molecular fingerprint of exhaled breath. This broad coverage enables capturing complex biochemical dynamics that reflect both systemic and localized respiratory physiology.

The instantaneous spectral bandwidth of the system presented here spans 2.9-11.5 µm (2580 $cm^{-1}$) in a single measurement. This exceeds the capabilities of current state-of-the-art broadband laser-based spectroscopy systems, including the recent demonstration by Liang et al. [12]. In that work, a maximum spectral coverage of 1010 $cm^{-1}$ was achieved by combining measurements from two distinct optical parametric oscillator-based laser systems, with the instantaneous bandwidth of a single source limited to 180 $cm^{-1}$. By contrast, our system achieves substantially greater instantaneous bandwidth – and thus access to a larger number of molecular species – while maintaining considerably lower experimental complexity. At the same time, the unparalleled sensitivity reported in their approach highlights the potential for continued advances in ultra-broadband laser-based infrared spectroscopy in breath analysis.

We demonstrate the ability to focus on specific compounds relevant to a given intervention (e.g., ammonia for protein intake, acetone for fasting, CO for smoking), while simultaneously capturing a broad range of other endogenous and exogenous molecules, highlighting the analyzer's multi-species measurement capability. The platform's high spectral resolution (0.1 $cm^{-1}$) allows for the detection of individual metabolic dynamics. Intervention-related dynamics, such as ammonia rise following protein intake, fasting-induced acetone changes, and smoking-related CO washout, illustrate the platform's ability to monitor both immediate and longitudinal metabolic responses.

A key strength of the platform is its standardized breath sampling system that permits the selection of specific phases of the exhalation. By continuously monitoring carbon dioxide, the end-tidal fraction of each exhalation was computed, allowing retrieved compound concentrations to be normalized accordingly. Although we focused on the collection of end-tidal breath in this study, the sampling system is adaptable and enables the targeted acquisition of any desired exhalation phase depending on the research objective. The platform demonstrated robust capability for online monitoring of intervention-related compounds, including acetone, $NH_3$, and CO, with high detection precision, and its performance was validated through external comparison with PTR-ToF-MS.

The information-rich absorbance spectra enable a variety of analytical strategies to fully leverage the spectral information. As demonstrated, the retrieved $NH_3$ concentration was optimized using normalization to the $CO_2$ concentration to account for sampling variations, while the acetone concentration was refined using relative absorbance spectra to minimize the effect of overlapping water-vapor absorption. Since the platform relies on the direct observation of a physically well-understood phenomenon like rovibrational absorption, it enables inherently calibration-free analyses, highlighting the system's reliability for absolute quantification without the need for chemical standards. Consequently, versatile data-processing strategies can be employed, for which machine learning-based approaches are expected to automate and enhance the extraction of information from the complex spectra.

In the context of molecular sensing, broadband infrared spectroscopy provides a complementary approach to mass spectrometry, with each method offering distinct advantages for the detection of specific classes of compounds [4]. Our infrared IDFG-FTS based platform uniquely provides direct access to small inorganic and volatile organic molecules such as CO, NO, $CH_4$, and $N_2O$, which are often challenging to detect by standard mass spectrometric techniques due to their low mass, high volatility, or poor ionization efficiency. Furthermore, these compounds are well characterized in terms of biochemical pathways or are informative exogenous markers of environmental or occupational exposure (e.g. CO, $N_2O$). The platform can quantify both exogenous compounds and endogenous metabolites within a single, calibration-free analytical framework, positioning it as a powerful platform for applications ranging from clinical to environmental health monitoring.

Our current system represents an initial step towards a versatile platform for breath analysis. While it already achieves broad spectral coverage, further extension of the spectrum – feasible with the present architecture [32] – would allow detection of additional compounds in the fingerprint region up to 20 µm. Enhanced sensitivity could enable the detection of trace compounds at concentrations below 1 ppbv, such as ethylene, and enable isotopologically resolved studies [12]. However, as both spectral coverage and sensitivity expand, the retrieval of molecular concentrations will become increasingly complex due to the high information density of the spectra. To address this challenge, integration of artificial intelligence and machine learning approaches is anticipated, with ongoing research focusing on optimized workflows for model training and evaluation [17].

Looking ahead, the potential of the ultra-broadband IDFG-based laser source to operate as a frequency comb suggests a pathway towards dual-frequency comb spectroscopy. Ultimately, this ultra-broadband platform provides a robust basis for the next generation of non-invasive, fast, real-time breath measurements, opening the door to time-resolved breath-to-breath molecular profiling in personalized clinical care.

## Methods

### *Experimental Design*

We developed a novel, compact spectroscopic breath platform based on ultrabroadband infrared laser spectroscopy. The system integrates three components: a breath sampling interface, an intrapulse difference-frequency generation (IDFG)-based ultrabroadband mid-infrared source, and a custom-built Fourier transform spectrometer (FTS) (Fig. 1).

### *IDFG-based Fourier transform spectroscopy*

An IDFG-based infrared supercontinuum source and a home-built FTS with turn-key operation were used for spectroscopic investigations in the spectral region 3 – 11.5 µm. The spectroscopic setup has been described in detail in previous work [14,47]. Therefore, we restrict our description to the most essential elements.

The light source was built on a Cr:ZnS ultrafast mid-infrared laser platform (CLPF-SC, IPG Photonics) [33,34], producing 3 W low-noise supercontinuum radiation spanning 2 – 11.5 µm, including ~300 mW in the 7 – 11.5 µm wavelength range at 80 MHz repetition rate. The source comprises a mode-locked master oscillator, a single-pass power amplifier (both pumped by an erbium-doped fiber laser), and an IDFG stage with a zinc germanium phosphide crystal. The output of this source effectively consists of two co-propagating components: an IDFG beam (4 – 11.5 µm) and the redshifted fundamental beam (2 – 7 µm).

The beam was coupled into a multi-pass gas cell (MPC, 31.2 m optical interaction path length HC30 L/M-M02, Thorlabs) equipped with uncoated ZnSe windows (WW70530, Thorlabs) for broadband transmission. After interaction with the sample, the light was directed to the home-built FTS. This interferometer uses a zinc selenide (ZnSe) beam splitter (BSW711, Thorlabs) to split the intensity of the beam into two equal parts propagating to a separate retroreflector (HRR201-P01, Thorlabs) mounted back-to-back on a single translation stage (DDSM100/M, Thorlabs). The beams were recombined on the beam splitter and propagated to two photovoltaic detectors (PVI-4TE-10.6, Vigo Photonics) with optimal detectivity in the spectral region 2.9 – 11.5 µm, defining the wavelength range of this system. As the interferograms on both detectors have opposite phases, a balanced detection scheme was used, which effectively doubled the interferogram signal while removing common-mode noise [35]. The light beam of a He-Ne laser was sent in parallel to the IDFG beam, and its interferogram was detected by a separate photodetector (PDA8A2, Thorlabs) to calibrate the optical path difference continuously.

Scanning the stage over 2.5 cm yielded a 10 cm optical path difference, corresponding to 0.1 $cm^{-1}$ (3 GHz) spectral resolution. One spectral scan was acquired in ~1.9 s, with 100 scans averaged per measurement to reduce the white noise in the spectrum and enhance the sensitivity.

*Breath sampling*

The online sampling system enabled direct measurements of the breath samples and a precise selection of a targeted fraction of the exhalation. Participants exhaled at 50 mL/s through a mouthpiece connected to an open-end buffer tube. An interface (GSI, Loccioni) provides feedback to self-regulate the exhalation flow rate for measurement standardization and indicates the exhalation phase using the $CO_2$ concentration measured by capnography. The capnograph (Capnostat 5, Philips) located between the mouthpiece and buffer pipe was used to trigger a three-way valve (MTV-3-N1/4U-32, Takasago) once the $CO_2$ concentration exceeded 3.5%, diverting most of the end-tidal breath into the MPC. After 10 s of sampling, the MPC was sealed to prevent dilution with the environmental air. The cell (0.85 L, 1 bar) was filled with two exhalations, evacuated to <1 mbar between samples, and heated to 40 °C. All sampling lines in contact with breath samples were heated to 40 ℃ to prevent condensation of relevant compounds and were made of the inert materials polytetrafluoroethylene (PTFE) and perfluoroalkoxy alkane (PFA). During the fasting case study, the MPC was evacuated after acquisition and the content was transferred to a 3-L Tedlar® bag (MediSense B.V., Groningen) using a pump (KNF Laboport N86) for subsequent PTR-ToF-MS analysis.

*PTR-ToF-MS analysis*

Tedlar® bags containing the exhaled breath samples were placed in a Heratherm™ oven (Thermo Scientific) and incubated for at least 10 minutes at 40°C. Exhaled breath samples were subsequently analyzed on a PTR TOF 8000 mass spectrometer (Ionicon Analytik, Innsbruck, Austria). The temperature, voltage, and pressure in the drift tube were set to 80°C, 480 V, and 2.14 mbar, respectively, resulting in a reduced electric field of 121 Td. The valve of the Tedlar bag® was attached to the 1/16" PTFE heated mass spectrometer inlet (80°C) by a series of PFA straight union reducers. The inlet flow rate was set to 25 mL/min and data were acquired at a rate of 2 Hz for the duration of 2 minutes. The scan range was set to *m/z* 0-234. For calibration of the instrument response, a 1-ppm gas mixture (Linde Gas Benelux, Dieren, The Netherlands) was used containing 8 reference compounds (methanol, acetaldehyde, acetone, isoprene, benzene, toluene, xylene, α-pinene). All data were processed in MATLAB R2014b (The MathWorks, Inc., Natick, MA).

*Design of the case studies*

Three different case studies were conducted with eight healthy participants under ethics approval by Research Ethics Committee of the Faculty of Science (Radboud University, REC22013) and written informed consent. Interventions included:

- **Protein-rich meal study:** Participants fasted ≥12 h, provided a baseline sample before the intervention (at time = 0 h), consumed a protein shake (0.83 g/kg body weight, HiPRO), and gave hourly samples thereafter. No other food was consumed and only water intake was permitted during the experiment.
- **Fasting study:** Participants fasted overnight and provided samples between 14 - 22 h post-fasting. On a control day, reference samples were collected without dietary restrictions. Only water intake was permitted during fasting.

- **Smoking study:** Regular smokers abstained ≥10 h prior to baseline sampling. After the first measurement at time = 0 h, each consumed one Marlboro Gold cigarette. The first post-smoking sample was collected within 10 min. No dietary restrictions were imposed on the participants in this case study.

*Data acquisition and processing*

Interferograms of the IDFG source and the reference He-Ne laser were digitized using an analog-to-digital converter (NI PCI-6251, National Instruments). The data-processing workflow is schematically shown in Fig. 6. Initial processing of the IDFG interferogram, such as resampling the optical path difference using the reference He-Ne interferogram, obtaining the (power) spectrum through FFT, and averaging multiple scans, was performed in real-time in a LabVIEW environment. In post-processing, the power spectra were converted to absorbance spectra in which the molecules' concentration scales linearly with the absorbance features. Based on the analysis, untargeted or targeted, different additional processing steps were performed. All postprocessing scripts have been written in MATLAB and are accessible via a public repository.

A total of seventy-eight breath samples were collected from eight healthy volunteers across three intervention-based case studies. The processed spectra from all seventy-eight samples have a spectral coverage of 2.9–11.5 µm (3450–870 $cm^{-1}$) with a spectral resolution of 0.1 $cm^{-1}$ (3 GHz). To assess statistical precision, five consecutive measurements (~3 minutes each) were acquired per breath sample, enabling evaluation of reproducibility and measurement stability.

The spectra were analyzed in two approaches:

- **Untargeted analysis:** The power spectra of the breath samples ($I_S$) were converted to absorbance spectra ($A$) using the power spectra of dry nitrogen gas ($I_0$) as background: $A = \ln{(I_0/I_S)}$. The absorbance spectra were subsequently baseline-corrected (asymmetric least squares, $\lambda = 1\times10^4$, $p = 0.001$), intensity-normalized on the absorption lines of $CO_2$, filtered for saturated absorption features of water based on HITRAN2020 [18] simulations, mean-centered. The postprocessed absorbance spectra were analyzed by principal component analysis (PCA, SVD-based) to reveal clustering of the breath samples by intervention, without retrieving the concentration of specified compounds.

- **Targeted analysis:** The absorbance spectra ($A$) were processed in two ways, either as absolute-concentration absorbance spectra ($A_A$) or relative-concentration absorbance spectra ($A_R$). The absolute-concentration absorbance spectra used the power spectra of dry nitrogen gas ($I_0$) as background: $A_A = \ln(I_0/I_S)$. The relative-concentration absorbance spectra used the power spectra of a breath sample from the same participant prior to the intervention ($I_{t_0}$) as background: $A_R = \ln(I_{t_0}/I_S)$. Therefore, $A_R$ exclusively contains absorbance features relating to the change in breath-sample composition. The concentrations of the compounds of interest were derived by least-squares fitting of reference absorbance spectra (HITRAN2020 or PNNL database) to the breath samples' absorbance spectra. The least-squares fit was complemented with polynomial baseline corrections (5th–8th order, depending on window) and the absorbance spectra were filtered for saturated absorption features of water. Multiple fits were applied to separate spectral windows to simultaneously retrieve the concentrations of several gases from the same breath sample. The least-squares models were optimized and validated *a priori* on a hybrid dataset to ensure accurate concentration retrieval [17].

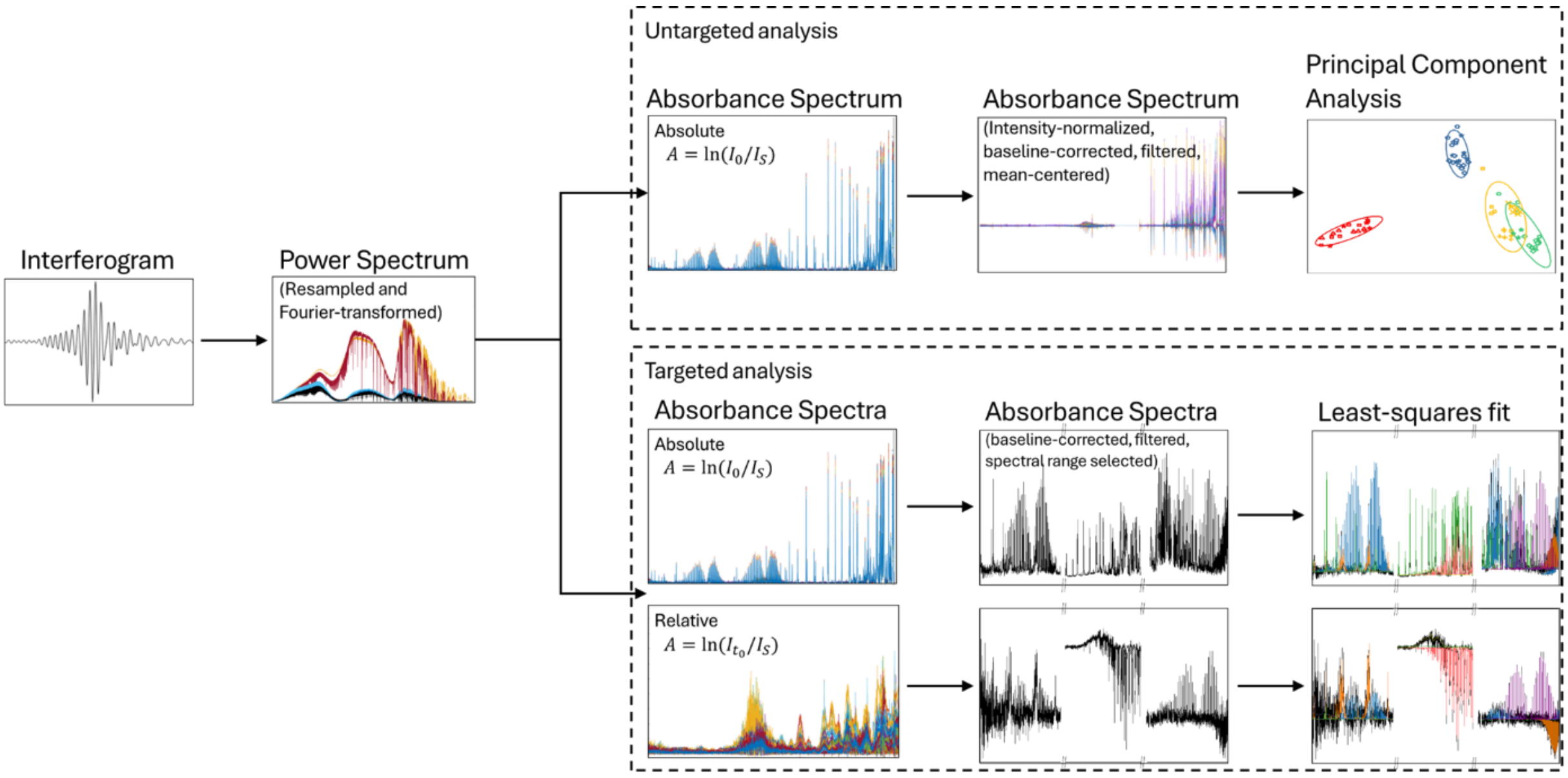

**Fig. 6. Data-processing workflow.** Acquired interferograms are converted into power spectra and subsequently into absorbance spectra, which are analyzed using principal component analysis to cluster breath samples according to interventions. A least-squares fit is performed to retrieve the concentrations of the compounds in each sample.

*Statistical Analysis*

Instrument precision was assessed by repeating the measurement of each breath sample five times. Because the composition of the sample in the MPC remained stable, the variance (± 1 standard deviation) between repeated measurements reflects the precision of the instrument and is represented as error bars in Fig. 5.

The measurement precision is compound-specific, depending on the absorption strength, absorption profile, spectral window (as the IDFG source has a non-uniform spectral density), and potential spectral overlap with other compounds. Accordingly, precision was estimated individually for each sample based on the variance across its five repeated measurements.

## Data availability

All data needed to evaluate the conclusions in the paper are present in the paper and/or the Supplementary Information, or can be retrieved from the Radboud Data Repository: https://doi.org/10.34973/hewn-jv32.

## Funding

The authors disclose support for the research of this work from the EU Horizon2020 program (grant 101015825, TRIAGE Project), and the Interdisciplinary Research Platform (IRP) at the Faculty of Science of Radboud University (grant AI for broadband spectroscopy).

## Author contributions

Conceptualization: RK, KvK, AK, SMC

Data curation: RK, MH, KvK

Formal analysis: RK, MH

Funding acquisition: AK, SMC

Investigation: RK, KvK, JM

Methodology: RK, KvK, JM, AK, SMC

Project administration: RK, SMC

Resources: RK, KvK, JM

Software: RK, KvK, AK

Validation: RK, KvK, JM

Visualization: RK, MH

Supervision: AK, SMC

Writing—original draft: RK

Writing—review & editing: RK, MH, KvK, JM, AK, SMC

## Competing interests

Authors declare that they have no competing interests.

# Tables

**Table 1. Typical compounds in breath samples with associated physiological basis.** Their typical concentration range in breath of healthy persons is given, together with their origin.

| **Volatile organic compound** | **Concentration range** | **Physiological basis** | Reference |
|---|---|---|---|
| **Acetaldehyde** | 0-5 ppm | Ethanol metabolism | 36 |
| **Acetic acid** | 0-50 ppb | Fermentation of non-digestible carbohydrates by the gut bacteria | 37–39 |
| **Acetone** | 100-2000 ppb | Decarboxylation of acetoacetate and the dehydrogenation of isopropanol | 21,40–42 |
| **Ammonia** | 50-1000 ppb | Catabolism of proteins and amino acids | 21,43 |
| **Butyric acid** | 0-10 ppb | Fermentation of non-digestible carbohydrates by the gut bacteria | 38,39 |
| **Carbon dioxide** | 4-6 % | Cellular respiration | 44 |
| **Carbon monoxide** | Non-smoker < 2 ppm<br>Smoker 5-12 ppm | Degradation of free hemoglobin and cellular hemoproteins by heme oxygenase in blood | 31,44–46 |
| **Ethane** | 0-10 ppb | Lipid peroxidation | 47–49 |
| **Ethylene** | 0-100 ppb | Lipid peroxidation of poly- and monounsaturated fatty acids | 50–52 |
| **Ethanol** | 10-1000 ppb | Fermentation of carbohydrates by gut microbes | 53,54 |
| **Isoprene** | 10-600 ppb | Lipolytic cholesterol metabolism in the skeletal muscle | 26 |
| **Methane** | 2-20 ppm | Methanogenesis of gut microbiota | 55,56 |
| **Methanol** | 100-400 ppb | Methionine amino acid metabolism via S-adenosyl methionine or fermentation of methylated compounds by gut bacteria | 57,58 |
| **Nitric oxide** | 0-50 ppb | Oxidation of L-arginine by nitric oxide synthase | 59,60 |
| **Nitrous oxide** | 300-1600 ppb | Reduction of nitrate compounds by denitrifying bacteria in the gut and oral cavity | 61,62 |
| **Propionic acid** | 0-50 ppb | Fermentation of non-digestible carbohydrates by the gut bacteria | 38,39 |
| **Water** | 2-4 % | Cellular respiration | 44 |

**Table 2. Compounds detectable with the current platform, with detection limits and spectral range.**

| **Volatile organic compound** | **Spectral range used ($cm^{-1}$)** | **Detection limit in 190 seconds** |
|---|---|---|
| Acetone | 1320-1460 | 60 ppb |
| Ammonia | 890-1192 | 4 ppb |
| Carbon dioxide | 945-1079 | 140 ppm |
| Carbon monoxide | 2055-2220 | 16 ppb |
| Ethylene | 870-1014 | 12 ppb |
| Isoprene | 850-950 | 14 ppb |
| Methane | 1250-1360 | 27 ppb |
| Methanol | 970-1080 | 25 ppb |
| Nitrous oxide | 2160-2260 | 16 ppb |
| Water | 850-1300 | 400 ppm |